\documentclass[twoside]{article}
\usepackage{qic,epsfig}
\usepackage{comment}
\usepackage{graphicx}
\usepackage{multirow}
\usepackage[table]{xcolor}
\usepackage{amsmath, amsfonts}
\usepackage{algorithm}
\usepackage{enumitem}
\usepackage{algpseudocode}

\makeatletter
\newcommand{\mathclap}[1]{\text{\hbox to 0pt{\hss$\m@th#1$\hss}}}
\makeatother

\begin{document}
\setlength{\textheight}{8.0truein}    


\normalsize\textlineskip
\thispagestyle{empty}
\setcounter{page}{1}


\vspace*{0.88truein}

\alphfootnote

\fpage{1}

\centerline{\bf
FACTORIZATION OF EXCLUSIVE-SUM-OF-PRODUCT EXPRESSIONS
}
\centerline{\bf WITH RECTANGLE COVERING TO REDUCE QUANTUM CIRCUIT COST}
\vspace*{0.37truein}
\centerline{\footnotesize
Audrey Hou, Lucia Zhang, Ali Al-Bayaty, Marek Perkowski \footnote{audreyhou12@gmail.com, lrzhang@usc.edu, albayaty@pdx.edu, h8mp@pdx.edu}}
\vspace*{0.015truein}
\centerline{\footnotesize\it Department of Electrical and Computer Engineering, Portland State University}
\baselineskip=10pt
\centerline{\footnotesize\it 1900 SW Fourth Ave, Portland, OR 97201, USA}
\vspace*{0.225truein}


\abstracts{
The implementation of quantum circuits is currently very expensive, especially due to the usage of large Toffoli gates. Therefore, it is critical to optimize circuit costs by factoring expressions as they become more complex. In the proposed algorithms to factor ESOP expressions, each product term is converted into a cell in a 2D matrix, and optimal factored AND/EXOR solutions are determined using Disjoint and Even-Odd Rectangle covering methods. Two programs implementing these algorithms were written in Python, tested, and evaluated using well-known benchmarks. The results showed that both literal counts used in classical logic circuits as well as the Maslov cost used in quantum circuits are reduced by 20\%-95\% depending on expression size.
}{}{}

\vspace*{10pt}

\vspace*{3pt}
\communicate{to be filled by the Editorial}

\vspace*{1pt}\textlineskip    
\section{Introduction}  
Quantum computing has the ability to solve many problems in various fields faster than classical computing. It is quickly advancing and estimated to be practical within a decade, but cost remains the biggest challenge. Therefore, it is critical to optimize quantum circuit synthesis, which will improve computing capacity with limited numbers of qubits. Expressions involving the operators AND and EXOR are commonly used in quantum circuit synthesis, thus the goal of this paper is to introduce a comprehensive method to optimally factor Exclusive-Sum-Of-Products (ESOP) expressions \cite{exorcism}, especially for quantum circuits built from generalized $n$-bit Toffoli gates, inverters, and Feynman gates. Transforming a two-level form into a multi-level form through factorization is critical especially for quantum circuits, because it helps avoid large and expensive Toffoli gate products present in two-level circuits. After factorization, we calculate the cost of the expression in three different ways. We use the literal count for classical circuits, Maslov costs for quantum oracle circuits with mirrors, and also Maslov costs for quantum automata circuits without mirrors \cite{mirrors}. Thus, multilevel AND/OR circuits are used in practice and one of the most often used ways to obtain them in industrial EDA systems is the iterated factorization methods \cite{saul 1, saul 2, saul 3, rajski, tsai, brayton 1, brayton 2, MIS}. However, there is not much work published on multilevel AND/EXOR synthesis that is important for quantum circuits built from binary reversible gates.

\par Few researchers have developed methods and tools for factoring ESOP expressions. Several researchers developed tools for multi-level logic synthesis based on EXOR gates. Saul’s algorithm for the multi-level synthesis of Reed Muller representations is based on a simple factorizer without much optimization\cite{saul 1, saul 2, saul 3}. Chattopadhyay (et al.) created a tool for multilevel AND/EXOR design \cite{KGPMIN}, but they did not use factorization. Tran (et al.) considered factorization where the factored terms are only affine functions \cite{tran 1, tran 2}. An important advantage of classical AND/EXOR circuits is their high testability. Rajski and Vesudevamurthy showed that the factorization of AND/OR expressions preserves tests, which means that the factorized logic network can be tested with the same set of tests as the original network \cite{rajski}. Tsai and Marek-Sadowska showed that a very simple and restricted factorization of AND/EXOR expressions preserves some tests and is useful in realization of arithmetic functions \cite{tsai}. 
\par Brayton (et al.) solved the factorization problem for Sum-of-Products (SOP) expressions by using a rectangle covering technique to come up with optimal solutions \cite{brayton 1, brayton 2, MIS}, targeted towards classical gates. 
Overlapping rectangle techniques can also be used for EXOR-Sum-of-Products (ESOP) factorization. Kalay (et al.) extended the overlapping rectangle technique for ESOP expression factorization \cite{kalay}. In our paper, we extend upon the ideas presented by Kalay and Perkowski, where we define two rectangle covering methods with different time complexities for the factorization of ESOP functions targeted towards quantum gates. We created software that includes many variants in method and process. This is the first time that the rectangle covering approach has been used for multi-level quantum circuit synthesis. 
\par This paper is organized as follows. In Section \ref{sec:background}, the basic information for factoring ESOP expressions is given. Section \ref{sec:esop rectangle coverings} presents our method for more efficient factorization of ESOPs using rectangle covering methods. Section \ref{sec:algorithms} introduces the new algorithms that were programmed to implement rectangle covering. Section \ref{sec:experimental} explains the benchmark selection and evaluation metrics. Section \ref{sec:results} analyzes the cost improvement from factoring. Section \ref{sec:conclusions} discusses the conclusions and future directions.

\section{Background}\label{sec:background}
\subsection{Basic definitions}
A Boolean variable is a variable that has a value of either 0 or 1. A literal is any instance of a Boolean variable like $a$, or its complement $\overline{a}$, which uses the NOT inverter operator. A cube (also referred to as a product term) is a conjunction or product of $d$ literals, where $d$ is called the degree of the cube.  
\par An Exclusive-Sum-Of-Products (ESOP) expression is an EXOR sum of cubes. 
We will use $E$ to denote an ESOP expression composed of $n$ cubes, written as
\begin{equation}
    E=\bigoplus_{k=1}^n c_k,
    \label{eq:esop_def}
\end{equation}
where $\bigoplus$ is the EXOR summation and each $c_k$ is a cube.
\begin{ex}
    An ESOP expression with four variables and six cubes of degree 2 is
    \[  E=ac\oplus \bar{b}c\oplus ad\oplus \bar{b}d \oplus af\oplus \bar{b}f.\]
\end{ex}
An ESOP expression is the canonical flattened form of AND/EXOR expressions. Specific types of ESOP expressions include Positive Polarity Reed-Muller (PPRM), Fixed Polarity Reed-Muller (FPRM), Kronecker Reed-Muller (KRM), Generalized Reed-Muller (GRM), and other similar two-level AND/EXOR forms \cite{yang}. In this paper, we will encounter expressions of the form FPRM and PPRM, which are defined below.
\begin{definition} 
Given a polarity $p = (p_1, p_2, \dots, p_n)$ such that $p_i \in \{0, 1\}$, the binary Fixed Polarity Reed-Muller (FPRM) form of a single-output Boolean function $f(x_1, x_2, \dots, x_n)$ is a function of the form 
\begin{equation}
f(x_1, x_2, \dots, x_n) = a_0 \oplus a_1 x_1^{(p_1)} \oplus a_2 x_2^{(p_2)} 
\oplus \dots \oplus a_{2^n-1} \left( x_1^{(p_1)} x_2^{(p_2)} \dots x_n^{(p_n)} \right),
\label{eq:fprm_form}
\end{equation}
with coefficients satisfying $a_i \in \{0, 1\}$, and literals defined by  
\[
x_i^{(p_i)} = 
\begin{cases}
\bar{x_i} & \text{if } p_i = 0, \\ 
{x}_i & \text{if } p_i = 1. 
\end{cases}
\]

The Positive Polarity Reed-Muller (PPRM) form is a special case of the FPRM expansion where all the literals are positive, meaning $p_i = 1$ for all $i$. 

\end{definition}
\par A general AND/EXOR expression is any expression involving Boolean literals and the operations AND, EXOR, and NOT. An AND/EXOR expression that is not an ESOP is often a factored ESOP. Given an ESOP expression $E$, we denote an equivalent factored representation as $F(E)$, or simply $F$ when the associated ESOP is clear from context. Unlike $E$, which is a two-level expression, $F(E)$ contains nested operations and multiple levels of AND and EXOR, resulting in a hierarchical structure.
Similar to ESOP, $F$ can also be written as an EXOR summation,
\begin{equation}
    F = \bigoplus_{k=1}^n f_k
\end{equation}
where each $f_k$ 
is a product of smaller AND/EXOR expressions. 
These smaller expressions can potentially be recursively factored to produce more nested levels. 
\begin{ex}
    An AND/EXOR expression that is not in ESOP expression form is:
    \[F=(ab\oplus c)(d\oplus e)\oplus bc,\]
    with $f_1=(ab\oplus c)(d\oplus e)$ and $f_2 = bc$. 
\end{ex}
\noindent 

\subsection{Boolean algebraic factorization}
A factored form of an ESOP expression $E$ gives an equivalent but more efficient representation, often containing fewer literals (counting repetitions) and fewer Boolean operators than the original ESOP, reducing the cost of the expression. 
\par Factorization of an ESOP expression is based on the algebraic rule, $ ab \oplus ac = a(b \oplus c)$, where $a$, $b$, and $c$ represent variables. 
The variables in expressions can be of two types: 
\begin{enumerate}
    \item Primary input variables, which are the literals of the Boolean expression. 
    \item Intermediate input variables that store expressions of some primary input variables and other intermediate variables. In some cases regarding quantum circuits, intermediate input variables must be ultimately restored, at some optimized point inside the quantum circuit, to the original primary input variables. This is done by mirroring. This is related to quantum cost and the number of ancilla qubits in reversible logic realization of these expressions, which will be discussed 
    in Sections~\ref{sec:higher degree} and \ref{sec:cost function}. 
\end{enumerate}
A simple algebraic factorization of an ESOP expression can be obtained by factoring out common literals, cubes, or subexpressions.
\begin{ex}
For example, given the ESOP expression,
\begin{equation*}
    E=ac\oplus bc\oplus ad\oplus bd \oplus af\oplus bf    
\end{equation*}
\indent A possible algebraic factorization is:
\begin{equation*}
F=c\left(a\oplus b\right)\oplus d\left(a\oplus b\right)\oplus f\left(a\oplus b\right)=\left(c\oplus d\oplus f\right)\left(a\oplus b\right)
\end{equation*}
\end{ex}
\par In some cases, axioms of Boolean logic can be incorporated into simple 
algebraic factorization to obtain better factorizations. The identity law allows the following algebraic reductions based on standard idempotent, complement, and inversion properties for AND and EXOR:
\begin{align}
a \oplus 1 &= \bar{a},\label{exor complement} \\
a\oplus a &= 0\label{exor idempotent}\\
a\oplus \bar{a} &= 1\\
a\cdot a &= a\label{and idempotent}\\
a\cdot \bar{a} &= 0
\end{align}
Below we illustrate an example that utilizes one of these Boolean laws.
\begin{ex}
    The ESOP expression $E =ab\oplus abcd$ can be factored as $F=ab(1\oplus cd)$. We can apply Eq.~\ref{exor complement} to get \[ab(1\oplus cd)=ab\overline{cd}.\]
\end{ex}
\subsection{Cost analysis quantities}
ESOP expressions are used in many types of circuits; the three that we focus on are classical binary circuits, quantum automata and quantum finite state machines \cite{kumar}, and quantum oracles \cite{henderson}. Therefore we use three variants of costs for comparing the circuit costs of an expression before and after factorization for these three uses. The literal count, which counts how many times any literal appears or is used in the expression, is used for determining the cost of the expression in a classical circuit. On the other hand, both the costs for quantum automata circuits and quantum oracle circuits are based on the Maslov cost (also called quantum cost) \cite{maslov}, which is a popular quantum cost function introduced by Dmitri Maslov that determines the cost of implementing quantum gates used in the quantum circuit realization of an expression. The difference is that automata circuits do not use mirrors, while oracle circuits require them.  
\par To evaluate the costs of a given AND/EXOR expression $E$, we will use the notation 
\begin{equation}
    \kappa(E)=(L(E), A(E), O(E)),
\end{equation} where $L(E)$, $A(E)$, and $O(E)$ represent the literal count for classical circuits, quantum automata circuit cost, and quantum oracle circuit cost, respectively. The explicit mathematical calculations and gate metrics for these costs are detailed 
in Section~\ref{sec:cost_calculations}.
\subsection{Determining expression and reversible circuit costs}\label{sec:cost_calculations}
AND/EXOR expressions, whether ESOP or factored ESOP, are made of the operators EXOR, AND, and NOT. A single output reversible circuit has a qubit for each primary input variable, as well as ancilla qubits when necessary and an output qubit. 
The gates used to realize AND/EXOR expressions are the inverter (Pauli-X), CNOT (Controlled-NOT/Feynmann), and generalized $n$-bit Toffoli gates. In particular, an inverter corresponds to NOT, a CNOT gate to EXOR, a $3$-bit Toffoli gate to AND, and an $n$-bit Toffoli gate to a cube of degree $n-1$. These gates are shown below as their circuit realizations. 

\begin{figure} [htbp] 
\vspace*{13pt} 

\centerline{
  \begin{minipage}[b]{0.2\textwidth}
    \centerline{\includegraphics{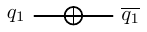}}
    \vspace*{4pt}
    \centerline{\footnotesize (a) Inverter}
  \end{minipage}
  \hfill
  \begin{minipage}[b]{0.22\textwidth}
    \centerline{\includegraphics{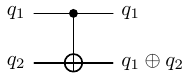}}
    \vspace*{4pt}
    \centerline{\footnotesize (b) CNOT}
  \end{minipage}
  \hfill
  \begin{minipage}[b]{0.24\textwidth}
    \centerline{\includegraphics{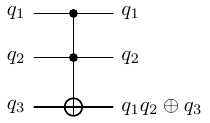}}
    \vspace*{4pt}
    \centerline{\footnotesize (c) Toffoli}
  \end{minipage}
  \hfill
  \begin{minipage}[b]{0.26\textwidth}
    \centerline{\includegraphics{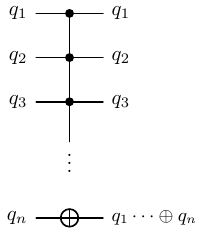}}
    \vspace*{4pt}
    \centerline{\footnotesize (d) Multi-Controlled}
  \end{minipage}
}

\vspace*{13pt} 
\fcaption{\label{fig:gates} Permutative Quantum Gates}
\end{figure}
The Maslov costs of the first few gate sizes is given below in Table \ref{tab:maslov}. The full table is shown in \cite{maslov}. In general, for a gate with $m$ controls, we will use the notation 
\[
    M(m+1) = \text{Maslov cost of an } (m+1)\text{-bit Toffoli}.
\]
\begin{table}[htbp]
    \tcaption{Maslov cost for generalized $(m+1)$-bit Toffoli Gate with $m$ controls and 0 ancilla qubits}
    \vspace*{10pt}
    \centerline{\begin{tabular}{|c|c|}
        \hline
        Gate qubit size ($m+1$) & Cost \\
        \hline
        1 (inverter) & 1 \\
        2 (CNOT) & 1 \\
        3 & 5 \\
        4 & 13 \\
        5 & 29 \\
        6 & 61 \\
        7 & 125 \\
        8 & 253 \\
        9 & 509 \\
        10 & 1021 \\
        $m+1 > 10$ & $2^{m+1}-3$\\
        \hline
    \end{tabular}}
    \label{tab:maslov}
\end{table}
\par To realize an AND/EXOR expression as a circuit, we use various gates to obtain the desired output value on the single output qubit, which is initialized with input value 0. We start by creating the values on the innermost levels, and then working outwards. The creation of the quantum circuit is slightly different between automata and oracle circuits, which will be discussed below. 
\par In quantum automata circuits, it is not necessary that the qubits be restored to the original values that were inputted, as all other qubits besides the target may be discarded as garbage/ancilla qubits. Therefore, once the desired output expression is obtained in the output qubit, no mirroring is needed. 
\par On the other hand, in quantum oracle circuits, the qubits must be restored to the original values that were inputted. Therefore, quantum gates that change the value of or have targets on qubits other than the output qubit must be \textit{mirrored}, which means that the gate must be reversed after the output value is obtained in order to restore the original input and undo the action on the target qubit.
\par
\begin{figure}[htbp]
    \vspace*{13pt}
    \begin{minipage}{0.45\textwidth}
        \centerline{\includegraphics{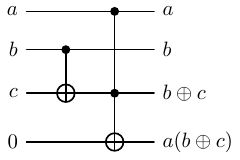}}
        \vspace*{4pt}
        \centering{\footnotesize (a) An example of a circuit used as a part of a Quantum Automata (no mirroring needed)}
    \end{minipage}
    \hfill
    \begin{minipage}{0.45\textwidth}
        \centerline{\includegraphics{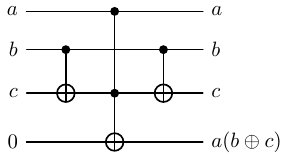}}
        \vspace*{4pt}
        \centering{\footnotesize (b) An example of a circuit used as part of a Quantum Oracle (mirrors needed)}
    \end{minipage}
    \vspace*{13pt}
    \fcaption{Realization of $a(b\oplus c)$ in Automata vs Oracle circuit\label{fig:automataoracle}}
\end{figure}

 As an example, to realize the expression $a(b\oplus c)$, we first look at the innermost levels $b$ and $c$. Because they are the default inputs, we look at the next level, $b\oplus c$, which can be created using a CNOT gate with control ${b}$ and target ${c}$. This makes the value of the ${c}$ qubit ${b\oplus c}$. Then, by using a 3-bit Toffoli, we can conjunct ${a}$ and ${b\oplus c}$ together to get the desired expression on the output line. This creates the automata circuit, as shown in Fig.~\ref{fig:automataoracle}a. Notice that the ${c}$ qubit is not restored, so in order to create the oracle circuit, we must use the same CNOT gate that we used first to restore the value of ${c}$, as shown in Fig.~\ref{fig:automataoracle}b.
\par The quantum cost of a final circuit is found by summing up the quantum cost of all gates used. To put it all together, we will use an example to summarize the different costs.
\begin{ex}
Let us look at ESOP expression $E=ab\oplus ac\oplus ad\oplus bc\oplus bd$, which can be algebraically factored as $F=ab\oplus (a\oplus b)(c\oplus d)$. By counting the number of literal instances, including repetitions, we find that $L(E)=10$ and $L(F)=6$. The numbers written below the expressions enumerate each literal to determine the literal count.
\begin{equation*}
    E=\underset{1}a\underset{2}b\oplus \underset{3}a\underset{4}c\oplus \underset{5}a\underset{6}d\oplus \underset{7}b\underset{8}c\oplus \underset{9}b\underset{10}d   
\end{equation*}
\begin{equation*}
F=\underset{1}{a}\underset{2}{b}\oplus(\underset{3}a\oplus \underset{4}b)(\underset{5}c\oplus \underset{6}d)
\end{equation*}
Next, we can realize both $E$ and $F$ as quantum circuits, which will allow us to determine the Maslov cost of each circuit and calculate the improvement from factoring. 
Also notice how a Toffoli gate naturally creates an EXOR on the output line, so no additional EXORs are needed.

\begin{figure}[htbp]
    \begin{minipage}{0.475\linewidth}
        \centerline{\includegraphics[height=3cm]{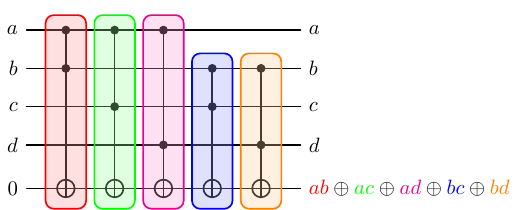}}
        \vspace*{4pt}
        \centering{\footnotesize (a) Circuit for expression before factorization}
    \end{minipage}
    \hfill
    \begin{minipage}{0.475\linewidth}
        \centerline{\includegraphics[height=3cm]{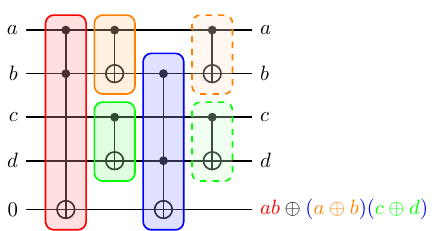}}
        \vspace*{4pt}
        \centering{\footnotesize (b) Circuit for expression after factorization}
    \end{minipage}
    \vspace*{13pt}
    \fcaption{Realization of E and F\label{fig:placeholder}}
\end{figure}
\par The quantum circuit for E uses five 3-bit Toffolis, giving a Maslov cost of $5\cdot M(3) = 25$, while the quantum circuit for F uses two 3-bit Toffolis and four CNOT gates, giving a Maslov cost of $2\cdot M(3)+4\cdot M(2) = 14$. In this example, the literal count improved by 4, and the Maslov cost improved by 11. As ESOP expressions grow larger, factorization will enable even greater cost improvements. 
\end{ex}

\begin{ex}
    Given the ESOP expression $E=ab\oplus a\overline{b}\oplus ac\oplus  cd\oplus \overline{b}d\oplus \overline{b}def\oplus cdef\oplus abdef\oplus abd$, a possible factored expression is 
    \begin{align*}
        F&=ab\oplus a\overline{b}\oplus ac\oplus \overline{b}d\oplus \overline{b}def\oplus  cd\oplus cdef\oplus abdef\oplus abd\\
        &= aab\oplus a\overline{b}\oplus ac\oplus \overline{b}d(1\oplus ef)\oplus cd(1\oplus ef)\oplus abd(1\oplus ef)\\
        &=a(ab\oplus \overline{b}\oplus c) \oplus (\overline{b}\oplus c\oplus ab)(d\overline{ef})\\
        &=(ab\oplus \overline{b}\oplus c)(a\oplus d\overline{ef})
    \end{align*}
    The realization of a circuit used in a quantum automata is depicted below, where no mirroring is required. The realization for a circuit used in a quantum oracle is similar, only all the gates besides the one highlighted in orange are mirrored across and used again to restore the affected qubits.
    \begin{figure}[htbp]
        \centerline{\includegraphics[width = 0.55\linewidth]{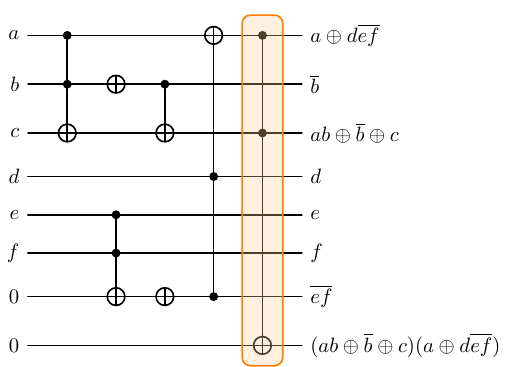}}
        \vspace*{13pt}
        \fcaption{Automata of After factoring}
    \end{figure}
    \par The cost of the automata circuit is 
    \[A(F)=M(3)+M(1)+M(2)+M(3)+M(1)+M(3)+M(3)=2\cdot 1+ 1\cdot 2+3\cdot 5=19,\]
    while for the oracle circuit, we must mirror each gate across the one highlighted in orange, which gives
    \[O(F)=2M(3)+2M(1)+2M(2)+2M(3)+2M(1)+2M(3)+M(3)=4\cdot 1+ 4\cdot 2+5\cdot 5=33.\]
    
\end{ex}

\section{ESOP Factorization Using Rectangle Covers}\label{sec:esop rectangle coverings}
In this section, two methods of rectangle covering for ESOP expression factorization are presented. First, we present the matrix representation of an ESOP expression. 
\subsection{Matrix representation of an ESOP expression}\label{sec:matrix rep}
Given an ESOP expression $E$ composed of $n$ cubes as described in Eq. \ref{eq:esop_def},
define the \textit{cube index set} to be
\begin{equation*}
C=\{1,2,\dots,n\},
\end{equation*}
where each $k\in C$ is the cube index of the cube $c_k.$ We can represent $E$ through a two-dimensional square \textit{matrix} of cells, where the nonzero values stored in the matrix correspond to indices of cubes in $E$. In order for the expression to be able to be converted to a matrix, the expression must be FPRM and the degree of each cube $c_k$ must be at most 2. Expressions that are not FPRM or with cubes of degree greater than 2 will be dealt with in Section~\ref{sec:higher degree}, where we describe two ways to convert higher degree expressions to degree 2. We will now discuss the conversion of $E$ to an associated matrix representation. 
\par Let $V=\text{supp}(E)$ be the support set of all variables used in $E$, and let $d$ be equal to the number of variables, $|V|$, where $||$ depicts the cardinality of the set. The expression $E$ can be represented by a square $d\times d$ matrix denoted $M$, where both the row and column labels consist of the elements of $V$, in alphabetical order for consistency. For each $r,c\in V$, we use $M_{r,c}$ to denote the number stored in the cell $(r,c)$ at row $r$ and column $c$, initialized as $M_{r,c}=0$ in an empty matrix.
\par Next, we map each cube to a cell in the matrix. Because $E$ has degree of at most 2, each of the cubes is either of the degree 2 form $ab$ or the degree 1 form $x$, where $a$, $b$, and $x$ can represent any variable in $V$. If a degree 2 cube $ab$ is has cube index $k$, we assign $M_{a,b}=M_{b,a}=k$. Note that although there are two symmetric instances of $k$ in $M$, both $M_{a,b}$ and $M_{b,a}$ refer to the same cube $ab$. If a degree 1 cube $x$, which can be rewritten as $xx$ by Eq.~\ref{and idempotent}, has cube index $k$, we assign $M_{x,x}$ gets numbered $i$. We now define a cell $(r,c)$ to be filled if $M_{r,c}\in \mathbb{N}$, and empty if $M_{r,c}=0$, which is implied by lack of a number in the matrix. 
\begin{ex}
To construct the matrix for the function $E = ab \oplus ac \oplus ad \oplus bc \oplus bd$, we first determine $V=\{a,b,c,d\}$, which gives a $4\times 4$ matrix. Then, we enumerate each  cube to obtain: $ab:1$, $ac:2$, $ad:3$, $bc:4$, $bd:5$, giving $C=\{1,2,3,4,5\}$. 
For each cube, we put the assigned number in the appropriate cell(s). As an example, for the cube $ab$, the number ``1" is placed in both cells $(a,b)$ and $(b,a)$, which is denoted as $M_{a,b}=M_{b,a}=1$. This matrix is shown below in Fig.~\ref{fig:matrix rep}.

\begin{figure}[htbp]
\centerline{\includegraphics{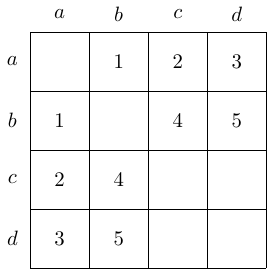}}
    \vspace*{13pt}
    \fcaption{Matrix Representation of $E$}\label{fig:matrix rep}
\end{figure}
\end{ex}
\subsection{Rectangles and solution sets}
In our matrix, we can ``cover" cells with \textit{rectangles} that correspond to factored expressions. We define these rectangle covers below.
\begin{definition}
Given two subsets $A,B\subseteq V$, a 
\textit{rectangle $R$} is defined as the ordered pair $R=(A,B)$, where $A$ and $B$ are called the factor sets of $R$. Think of them as the length and width dimensions, their Cartesian product determining the set of cells covered by the rectangle. 
Furthermore, the associated factored expression for the rectangle is a generalized cube,
\begin{equation}\label{eq: factored expression rectangle}
F_{R}=\left(\bigoplus_{a\in A} a\right)
\left(\bigoplus_{b\in B} b\right),
\end{equation}
where $\bigoplus$ is the EXOR summation. 
\end{definition}
\par To determine how useful a rectangle $R$ is, we define the density of $R$ as the percentage of cells it covers that are filled.
\begin{definition}
Let $C_R$ be the set of values/cube indices in the filled cells of $R$,
\[C_R=\{M_{r,c}\neq 0\mid r\in A, c\in B\},\]
Then, the density of a rectangle $R$ is defined as the percentage
\begin{equation}
\rho(R) = \left(\frac{|C_R|}{|A||B|}\right)\times 100.
\end{equation}
\end{definition}
\par We denote a rectangle by circling all of the cells that it covers and enclosing the circles with a border. In particular, filled cells will be enclosed with solid lines, while empty cells with dashed lines. Some examples of rectangles of different density are depicted below in Fig.~\ref{fig:rectangle density ex} on the matrix from Fig.~\ref{fig:matrix rep}.
\begin{figure}[htbp]
\centerline{\includegraphics[scale=0.9]{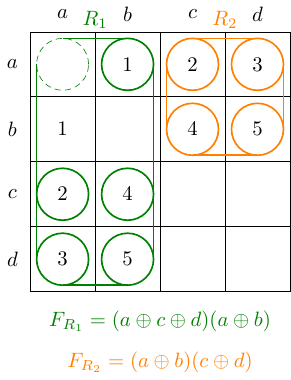}}
    \fcaption{Examples of Rectangles}\label{fig:rectangle density ex}
\end{figure}
\par Let the rectangle circled and enclosed in green be Rectangle $R_1$, which has $C_{R_1}=\{1,2,3,4,5\},$ and the one in orange be Rectangle $R_2$, which has $C_{R_2}=\{2,3,4,5\}$. We can see that $R_1=(\{a,c,d\}, \{a,b\})$ contains $|C_{R_1}|=5$ filled cells out of $3\cdot 2=6$ total circled cells, giving a density of $\frac{5}{6}=83.\overline{3}\%$. Notice that although cells $(b,a)$ and $(b,b)$ look like they are enclosed within the borders of $R_1$, they are not circled and therefore not part of the rectangle. We can also see that $R_2=(\{a,b\},\{c,d\})$ contains 4 filled cells out of 4 total covered cells, giving a density of $\frac{4}{4}=100\%$.
\par Now that we have the definition and properties of a rectangle, we can define a rectangle covering solution, which will allow us to find a factored expression of the original ESOP expression. The following two sections classify the types of rectangle coverings scenarios that we have, specifically regarding the numbers of rectangles that cover each cell. 
\subsection{Disjoint rectangle covering}\label{sec:disjoint}
A disjoint rectangle covering is a solution set $S$ of mutually disjoint rectangles 
such that every filled cell in the matrix $M$ is covered by exactly one rectangle in $S$, and every empty cell is covered by zero rectangles. Two rectangles are disjoint if they do not cover any common cells.
For disjoint rectangle covering, we will only use rectangles that have 100\% density, meaning all of the cells it covers are filled cells. 
Furthermore, the factored expression $F$ associated with the solution set $S$ is the EXOR sum of the factored expressions of each rectangle, and because every cube of $E$ is accounted for exactly once by the rectangles of $S$, we can conclude
\begin{equation}\label{eq:total factored expression}
F=\bigoplus_{R\in S}F_{R}=E.
\end{equation}
\begin{ex}
Using the matrix from Fig.~\ref{fig:matrix rep}, we can demonstrate examples of disjoint rectangle covering. In Fig.~\ref{fig:disjoint covering}, we show two solutions out of many possible coverings. 

\begin{figure}[htbp]
    \centerline{\begin{minipage}{0.4\linewidth}
        \centerline{\includegraphics{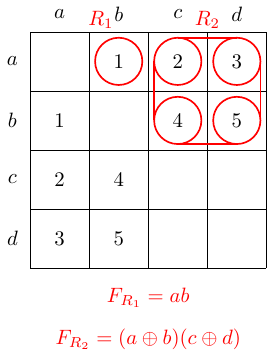}}
        \vspace*{4pt}
        \centering{\footnotesize (a) Solution 1}
    \end{minipage}
    \hspace{1cm}
    \begin{minipage}{0.4\linewidth}
        \centerline{\includegraphics{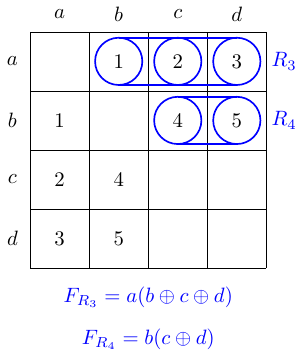}}
        \vspace*{4pt}
        \centering{\footnotesize (b) Solution 2}
    \end{minipage}
    \vspace*{13pt}}
    \fcaption{Disjoint Rectangle Covering Example}\label{fig:disjoint covering}
\end{figure}

\par \noindent For solution $S_1=\{R_1,R_2\}$ in red, we have $C_{R_1}=\{1\}$ and $C_{R_2}=\{2,3,4,5\}$, which means that all five filled cells in the matrix are accounted for exactly once. 
Therefore, the factored form is 
\[F_1 = (a\oplus b)(c\oplus d)\oplus ab,\]
and we find that $\kappa(F_1)=(6,12,14)$. For the solution $S_2=\{R_3,R_4\}$ in blue, we have $C_{R_3}=\{1,2,3\}$ and $C_{R_4}=\{4,5\}$. We obtain the factored form, 
\[F_2 = a(b \oplus c \oplus d) \oplus b(c \oplus d),\]
and we find that $\kappa(F_2)=(7,13,16)$, which is more expensive than $F_1$.
\end{ex}

\subsection{Even-odd rectangle covering}\label{sec:evenodd}
In disjoint rectangle covering, we are restricted to exactly 1 rectangle cover per cell. However, using the generalized concept of odd and even coverage allows us to obtain better solutions because we can utilize larger rectangle covers and thus have fewer numbers of them, reducing the number of Toffoli gates.
\begin{definition}
A solution set $S$ of rectangles on a matrix is a valid even-odd rectangle covering if every filled cell gets covered by an odd number of rectangles, and if every empty cell is covered by an even number of rectangles. This is because of the EXOR idempotent law in Eq. \ref{exor idempotent}, which allows us to say that if we have an filled cell with cube $x$, then covering $x$ an odd number of times preserves equality. 
\begin{equation}
E=E'\oplus x=E'\oplus x\oplus\underbrace{x\oplus\ldots\oplus x}_{\mathclap{\text{even number of additional existing cube}}}
\end{equation}
If we have an empty cell corresponding to cube $y$, then covering it an even number of times preserves equality.
\begin{equation}
E=E\oplus \underbrace{y\oplus\ldots\oplus y}_{\mathclap{\text{even number of additional non-existing cube}}}
\end{equation}
Similarly, the factored expression of the solution set $S$ is the same as stated in Eq. \ref{eq:total factored expression}. Note that this even-odd cell coverage technique in AND/EXOR expressions does not exist for AND/OR expressions. 
\end{definition}
\par The following example illustrates an even-odd rectangle covering utilizing odd coverage on one of the cells. 
\begin{ex}
\label{ex:odd}
    For the expression given below, where the numbers written below the cubes are the indices of $C$,
    \[E=\underset{1}{ac}\oplus \underset{2}{bc}\oplus \underset{3}{ad}\oplus \underset{4}{bd}\oplus \underset{5}{cd}\oplus \underset{6}{de}\oplus \underset{7}{ae}\oplus \underset{8}{ce}\oplus \underset{9}{af}\oplus \underset{10}{ef},\]
    the cost is $\kappa(E)= (20,50,50) $. To compare odd rectangle covering with disjoint rectangle covering, two solutions of covering the matrix for $E$ are shown in Fig.~\ref{fig:odd covering}.
\begin{figure}[htbp]
    \centerline{\begin{minipage}{0.475\linewidth}
        \centerline{\includegraphics[scale=0.8]{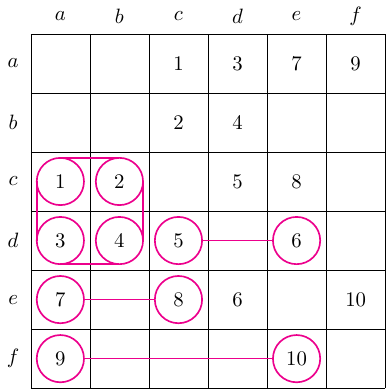}}
        \vspace*{4pt}
        \centering{\footnotesize (a) Disjoint Rectangle Covering}
    \end{minipage}
    \hfill
    \begin{minipage}{0.475\linewidth}
        \centerline{\includegraphics[scale=0.8]{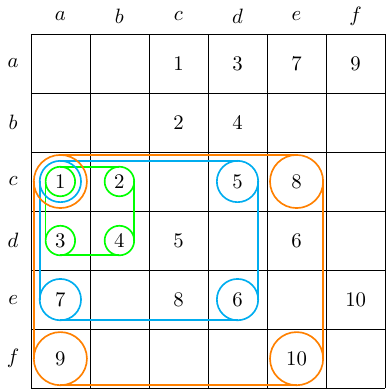}}
        \vspace*{4pt}
        \centering{\footnotesize (b) Solution with Odd Coverage on cell $M_{a,c}=1$}
    \end{minipage}}
    \vspace*{13pt}
    \fcaption{Odd Rectangle Covering vs. Disjoint Covering on the same matrix \label{fig:odd covering}}
\end{figure}

\par The factored form obtained without odd cell coverage using disjoint rectangle covering is 
\[F_1 = (a\oplus b)(c\oplus d)\oplus d(c\oplus e)\oplus e(a\oplus c)\oplus f(a\oplus e),\]
which has cost $\kappa(F_1)=(13,25,30)$. 
The factored form using odd cell coverage is 
\[F_2 = (a \oplus  b)(c \oplus  d) \oplus  (a \oplus  c)(d \oplus  e) \oplus  (a \oplus  e)(d \oplus  f),\] 
which has a better cost of $\kappa(F_2)=(12,21,27)$. 
\end{ex}
Similarly, we will demonstrate an example of even-odd rectangle covering that utilizes even coverage on one of the cells. 
\begin{ex}
    For the expression given below
    \[E=\underset{1}{bc}\oplus \underset{2}{ad}\oplus \underset{3}{bd}\oplus \underset{4}{ce}\oplus \underset{5}{af}\oplus \underset{6}{ef}\]
    two rectangle covering solutions are shown in Fig.~\ref{fig:even covering}, one disjoint and one even-odd.
\begin{figure}[htbp]
    \centerline{\begin{minipage}{0.475\linewidth}
        \centerline{\includegraphics[scale=0.7]{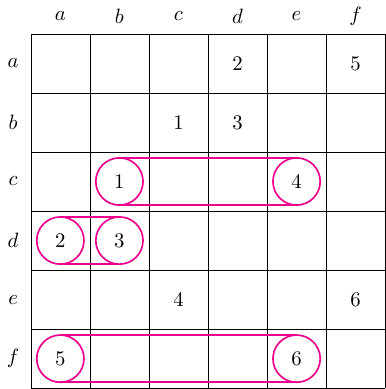}}
        \vspace*{4pt}
        \centering{\footnotesize (a) Disjoint Rectangle Covering}
    \end{minipage}
    \hfill
    \begin{minipage}{0.475\linewidth}
\centerline{\includegraphics[scale=0.7]{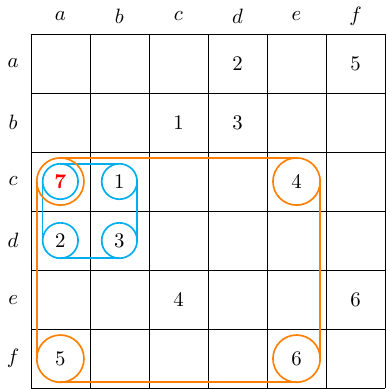}}
        \vspace*{4pt}
        \centering{\footnotesize (b) Solution with Even Coverage on $M_{a,c}$}
    \end{minipage}}
    \vspace*{13pt}
    \fcaption{Even Rectangle Covering vs. Disjoint Covering on the same matrix \label{fig:even covering}}
\end{figure}

\par The factored form obtained without even cell coverage using disjoint rectangle covering is: 
\[F_1 = c(b \oplus  e) \oplus  d(a \oplus  b) \oplus  f(a \oplus  e)\]
which has cost $\kappa(F_1)=(9,18,21)$.
\par In Fig. \ref{fig:even covering}b, the non-existing cube $ac$, denoted by bolded red ‘7’ in $M_{c,a}$, is introduced and covered by two rectangles, once by each of the blue and orange rectangles.  The optimal factored form with even coverage is:
\[F_2 = (a \oplus b)(c \oplus d) \oplus (c \oplus f)(a \oplus e)\]
which has a better cost $\kappa(F_2) = (8, 14,18)$.  

\end{ex}

\section{Algorithms}\label{sec:algorithms}
In this section we introduce algorithms that allow us to find optimal rectangle covering solutions. Given a matrix representation of an ESOP expression as described in Section~\ref{sec:matrix rep}, our algorithms are designed to search throughout the matrix for rectangles based on certain criteria. Specifically, we introduce two pseudocode algorithms, which have different goals and time complexities. The first one finds the quasi-optimal disjoint rectangle covering using a greedy method, while the other uses a recursive method that implements even-odd cell coverage to find the most optimal even-odd covering solution. This will allow us to compare the results between a heuristic and exhaustive search over specified sets of rectangles. 
\subsection{Disjoint rectangle covering algorithm}
The disjoint rectangle covering algorithm, abbreviated as Algorithm D, finds disjoint rectangle coverings using only 100\% density rectangles, as demonstrated in Section~\ref{sec:disjoint}. The goal for this algorithm is to find a quasi-optimal solution with polynomial time complexity, meaning that it may not produce the ultimately most optimal cost reduction, but it is a decently good reduction. \par This will be done using a greedy method that finds the largest (covering the most cells) 100\% density $2\times n$ rectangle in the matrix. Then, after adding that rectangle to the set of rectangles $S$, all the cells in that rectangle are emptied. This process is repeated until the matrix is completely empty. 

\begin{algorithm}[H]
    \caption{Find Disjoint Rectangles Optimally in Polynomial Time}
    \label{alg:d} 
    \begin{algorithmic} 
        \Require 2D matrix $M$
        \State largest rectangle $ R \gets [ \ ]$
        \State solution list $S \gets [\ ]$
        \State dimension $d \gets $ length of $matrix$
        \While{$M$ is not empty}
            \For{$r_1$ from 0 to $d$}
                \For{$r_2$ from $r_1+1$ to $d$}
                    \State current rectangle $R_c\gets [ \ ]$
                    \For{$c$ from 0 to $d$}
                        \If{$M[r_1][c]$ and $M[r_2][c]$ are filled}
                            \State add cell pair to $R_C$
                        \EndIf
                    \EndFor
                    \If{number of cells in $R_C>$ number of cells in $R$}
                        \State $R \gets R_C$
                    \EndIf
                \EndFor
            \EndFor
            \State add $R$ to $S$
            \State empty cells covered by $R$ in $M$
        \EndWhile
    \end{algorithmic}
\end{algorithm}
\subsection{Even-odd rectangle covering algorithm}
The Even-Odd Rectangle covering algorithm, abbreviated as Algorithm EO, finds even-odd rectangle coverings as described in Section~\ref{sec:evenodd}. The goal for this algorithm is to find the most optimal solution, meaning it has the least expensive quantum circuit cost. This will be implemented using a recursive traversing search through all rectangles in the matrix that satisfy certain criteria. The time complexity was not prioritized in the design because the main focus is to find the best solution, which ultimately made the search too long for some large examples.
\par In Algorithm D, the rectangles were limited to the form $2\times n$ in order to keep the run time polynomial. In Algorithm EO, to examine all possible rectangles, we run through all 
possible subset pairs $(A,B)$ where $A,B\subseteq V$. Given a feasibility bound constant, $F$, all rectangles containing more than 1 cell and a density of at least $F$ get added to a list of good rectangles, $G$. 
Then, for each rectangle in $G$, we perform a recursive method by removing that rectangle, finding new rectangles on the new matrix, and continuing the process until the matrix is empty. When removing a rectangle from the matrix, we empty all filled cells, however, all empty cells that the rectangle covers get filled with a new number not yet in $C$.

\begin{algorithm}[H]
    \caption{Find best solution using recursion and Even/Odd $Rectangle$ Covering}
    \label{alg:eo}
    \begin{algorithmic} 
        \Require matrix $M$, current solution $S$, current cell count $C$
        \If {$M$ is empty}
            \State \Return $[S]$
        \EndIf
        \State list of all solutions $A \gets []$
        \State list of possible rectangles $P \gets []$
        \For {each subset $A$ of $rows$ in matrix}
            \For {each subset $B$ of $columns$ in matrix}
                \If {$len(A)\times len(B)\geq2$}
                    \State rectangle $R \gets Rectangle(A, B, matrix)$
                    \If {$\rho(R) \geq F$}
                        \State add $R$ to $P$
                    \EndIf
                \EndIf
            \EndFor
        \EndFor
        \For {rectangle $R$ in P}
            \State $M' \gets$ deep copy of $M$
            \State new solution $S' \gets$ $S\cup [R] $
            \State $C'\gets C$        
            \For {cell in $R$}
                \If {cell is filled}
                    \State make cell empty
                \Else
                    \State $C'\gets C'+ 1$
                    \State fill cell with number $C'$
                \EndIf
            \EndFor
            \State add solutions from findRectangles($M', S', C'$) to $A$
        \EndFor
        \State \Return all solutions $A$
 
    \end{algorithmic}
\end{algorithm}
\subsection{Handling ESOPs with higher-degree cubes}\label{sec:higher degree}
Earlier, it was mentioned that the conversion of an ESOP expression to a matrix required an ESOP expression that was FPRM and had degree of at most 2. Now, we will discuss two methods to account for ESOPs with a higher degree or with mixed polarities by converting the expression such that it satisfies the requirements for the matrix conversion.
\par In an ESOP expression with mixed polarity for a variable, say $a$, the issue is that both the literals $a$ and $\bar{a}$ appear in the same expression. We can either treat $a$ and $\bar{a}$ as separate variables, or use the EXOR complement rule in Eq.~\ref{exor complement} to convert them to the same polarity, such that for each variable, only one polarity is used. This means that either all instances of $a$ get turned into $\overline{a}$ or vice versa. 
One way to solve this issue is to convert the expression to FPRM using the EXOR complement rule in Eq.~\ref{exor complement}, such that for each variable, only one polarity is used. This means that either all instances of $a$ get turned into $\overline{a}$, or vice versa, giving two options. To determine the polarity that produces the best FPRM form, we run all $2^{|V|}$ possible polarities, converting variables using that polarity, keeping track of costs, and then taking the polarity with the cheapest initial cost. 
Another way to solve this issue is to treat $a$ and $\bar{a}$ as separate variables, meaning we replace each negated instance of a variable with a new input variable that is not in $V$. We call these two versions the FP (fixed polarity) and DP (dual polarity) versions, respectively. Both methods are used in the algorithm, converting the expression to a usable expression that can be converted to a matrix.
\par After the expression has variables of fixed polarity, we may have cubes that have degree greater than 2. We need to make them degree 2, so we look for cubes that get repeated multiple times. For each cube, we use a ranked scoring system of frequency * degree, where frequency is the number of times it appears in or divides a term. We find the cube with the best score, substitute it with a new variable not in V, and continue this process of replacing cubes with new variables not in $V$ until all remaining cubes have degree 2 or less. Through the two polarity processes, we yield two canonically equivalent but different representations of the same higher degree expression, but with maximum degree 2. We input both of these expressions into our method pipeline. 
\par Once the matrix pipeline finishes, we get a factored form. Then, we must substitute back in our variables. If the frequency of the substitution was only 1, then we substitute back in the cube that it replaced. However, if the frequency of the substitution was more than 1, then we keep the substituted variable, and then create an ancilla for that substitution because it will help reduce reusing the same Toffoli gate multiple times. The calculation of the cost will be discussed in the next subsection.
\begin{ex}
    Given the ESOP expression \[E=bd\oplus d\oplus abcd\oplus acd\oplus abcde\oplus bcde\oplus be\oplus b\oplus abef,\] the algorithm gives the substitutions $g=acd, h=be,i=cd,j=af$, yielding the 2D expression \[E_m=bd\oplus d\oplus bg\oplus g\oplus gh\oplus hi\oplus h\oplus b\oplus hj.\] Putting this through Algorithm D gives the factored expression
    \[F_m=(g\oplus i\oplus j\oplus 1)(h)\oplus (b)(d\oplus g\oplus 1)\oplus d\oplus g.\]
    We see that g appears 3 times, while the other three substitutions each only appear once. Thus, we use an ancilla to store the value of $g=acd$ but we substitute the other three substitutions back into the expression to get
    \[F=(g\oplus cd\oplus af\oplus 1)(be)\oplus (b)(d\oplus g\oplus 1)\oplus d\oplus g.\]
    Now, in this next section, we will demonstrate a model for calculating the final cost of a final expression without realizing it as a circuit. 
\end{ex}

\subsection{A recursive Maslov-based realization model for calculating quantum cost}\label{sec:cost function}
In many cases, the quantum cost of our expressions can be calculated by breaking apart the expression and figuring how many and which gates are used, rather than by going through the process of realizing a circuit. Thus, we create a Maslov-based realization model targeting the form of our expressions. It is a consistent algorithm that is helpful for bulk cost calculation with a large number of benchmarks.  
\par For an unfactored ESOP expression $E$ with $n$ cubes, each cube $c_k$ with degree $d$ corresponds to a $(d+1)$-bit Toffoli gate with $d$ controls and a target on the output line. For each negated variable, we will assume that an inverter is used both before and after its usage in a gate in order to return the value of the input qubit back to its original polarity. There may be instances where two inverters may cancel out, but because inverters are cheap, we will consistently count 2 for each negation. 
This means that the costs can be denoted as 
\[A(c_k)= O(c_k)=M(d+1)+2\cdot \text{\# of inverters} \cdot M(1).\]
Therefore, the total cost of the expression can be found by summing the costs of each term: 
\[A(E) = \sum_{k=1}^n A(c_k) \quad \text{and} \quad O(E) = \sum_{k=1}^n O(c_k)\]

\par On the other hand, it is more difficult to calculate the exact minimum cost of a factorized expression, because there are many possible ways to realize an expression. Therefore, we define a systematic realization model that helps estimate the quantum cost of an expression. 

\par Given the factored expression $F$ and list of remaining substitutions with frequency greater than 1 from Section~\ref{sec:higher degree}, each substitution gets its own intermediate ancilla qubit to store the value of the substituted cube. For the automata cost, the cost of computation is counted once, but for the oracle cost, each ancilla has to restored back to its input value so the cost of computation is doubled. Recall Eq. \ref{eq:total factored expression}, which states that the factored expression $F$ is the EXOR sum of all $F_R$ such that $R\in S$. This means that
\[A(F) = \sum_{R\in S} A(F_R) \quad \text{and} \quad O(F) = \sum_{R\in S}^n O(F_R)\]
Also recall Eq. \ref{eq: factored expression rectangle}, which states that $F_R$ is of the form $E_1E_2$
where $E_1=\bigoplus_{a\in A} a$ and $E_2=\bigoplus_{b\in B} b$ are both ESOP expressions. To calculate the cost of $F_R$, we see that we must use a 3-bit Toffoli gate with a control on a qubit with the value of $E_1$ and a control on a qubit with the value of $E_2$, which gives 
\[A(F_R)=M(3)+A(E_1)+A(E_2) \quad \text{and}\quad O(F_R)=2\cdot M(3)+O(E_1)+O(E_2) \]
There is a special case when one of the ESOP subexpressions, say $E_1$, is a single cube with degree $d_1$. In this case, we do not want to realize $E_1$ on a new ancilla because it will not be used again (if it were used again, we wouldn't have substituted it back in earlier section), which would mean that $A(f_k)=M(2+d_1)+A(E_2)$. Furthermore, if $E_2$ is also a single cube with degree $d_2$, then $A(f_k)=M(2+d_1+d_2)$. For oracle costs, the formulas are the same, replacing the function $A$ with $O$. 
\par Otherwise, when either factor $E_1$ or $E_2$ is an ESOP of multiple cubes, its value is realized on a dedicated ancilla qubit, which is approached differently depending on circuit type. 
The automata cost model allocates a new ancilla per factor evaluation without reuse of intermediate values, while the oracle cost model allows reuse of ancillas across factors due to uncomputation. Specifically, at most two ancillas are required for oracle evaluation, one for $E_1$ and another for $E_2$, so the two ancillas can be reused after uncomputation for each rectangle expression.
An example showing how the cost function is shown below. 

\begin{ex}
    Continuing the previous example, we had a factored expression 
    \[F=(g\oplus cd\oplus af\oplus 1)(be)\oplus (b)(d\oplus g\oplus 1)\oplus d\oplus g.\]
    where $g=acd$ was a variable that appeared 3 times, meaning that $g$ gets its own ancilla intermediate variable. We must add the cost of $g$ at the end. 
    \par We consider each of the expressions associated with the four rectangles. The first, $F_{R_1}=(g\oplus cd\oplus af\oplus 1)(be)$, is one of the special cases because $E_2$ is a single cube. For calculating the ancilla costs, because $g$ is used elsewhere in the expression, we use a separate ancilla to store the value for $E_2=g\oplus cd\oplus af\oplus 1$. Thus, we get 
    \begin{align*}
    A(F_{R_1})&=M(2+2)+A(g\oplus cd\oplus af\oplus 1)\\
    &=M(4)+M(2)+M(3)+M(3)+M(1)=25
    \end{align*}
    For calculating the oracle costs, due to mirroring we can store the value for $E_2$ directly on $g$ by computing $cd$ and $af$ onto the qubit, using one less ancilla than the automata circuit. Thus, we get
    \begin{align*}
        O(F_{R_2}) &= M (2+2)+O(g\oplus cd\oplus af\oplus 1)\\
        &= M(4)+2M(3)+2M(3)+2M(1) = 35
    \end{align*}
    We repeat this process for the remaining three rectangles to get
    \[A(F)=25+7+1+1+A(acd)=47 \quad \text{and} \quad O(F) = 35+9+1+1+O(acd)=72.\] 
\end{ex}

\section{Experimental Setup}\label{sec:experimental}
\subsection{Benchmark selection}
To compare the two rectangle covering algorithms and evaluate the cost improvements, benchmarks were primarily drawn from standard PLA \cite{pla} format benchmark collections derived from the MCNC benchmark suite \cite{MCNC}. 
PLA files were converted to minimized ESOP forms using EXORCISM-4 \cite{exorcism}, a popular logic synthesis tool for minimizing ESOP expressions. Then, using the minterms in the ESOP files, the FP and DP variants ESOP expressions were generated. Benchmarks that could not be run in a feasible amount time were excluded.

\subsection{Evaluation metrics}\label{sec:evaluation metrics}
To evaluate the effectiveness of the proposed methods, the following metrics were used:
\begin{itemize}[noitemsep]
    \item $L_E$: Initial literal count of the ESOP expression, equivalent to the literal count of the ESOP after EXORCISM-4.
    \item $L_F$: Final literal count after factorization and rectangle covering.
    \item $\Delta L = \frac{L_E - L_F}{L_E}\times 100\%$: Percentage literal count reduction.
    \item $A_E = O_E$: Initial automata/oracle (equivalent) cost computed from the initial ESOP representation.
    \item $A_F$: Final automata cost after synthesis.
    \item Anc.$_A$: Number of ancillas used in the automata circuit
    \item $\Delta A =\frac{A_E - A_F}{A_E}\times 100\%$: Percentage Automata cost reduction.
    \item $O_F$: Final oracle cost after synthesis.
    \item Anc.$_O$: Number of ancillas used in the oracle circuit
    \item $\Delta O = \frac{O_E - O_F}{O_E}\times 100\%$: Percentage Oracle cost reduction.
\end{itemize}

\subsection{Experimental procedure and pipeline}
Below is the full procedure for factorizing a given ESOP expression into an optimally factored AND/EXOR expression using our two algorithms.
\begin{enumerate}[noitemsep]
\item Obtain benchmark ESOP expression $E$.
\item Convert $E$ to a degree-2 fixed or dual polarity representation.
\item Construct the corresponding matrix representation.
\item Compute the initial literal count and realization costs.
\item Apply Algorithm D and Algorithm EO (with feasibility bound 0.9) independently.
\item Convert each rectangle covering solution into a factored expression.
\item Compute and record the final metrics.
\end{enumerate}

All costs are computed using the Maslov cost model defined in Section~\ref{sec:cost function}. 
Comparisons were performed between Algorithm D and Algorithm EO, as well as between Fixed-Polarity and Dual-Polarity methods of ESOP file decomposition and processing. 

    
\section{Results}\label{sec:results}
\subsection{Tables and Figures}
\par Table \ref{tab:100} illustrates the improvement for each benchmark that was run. In addition, the number of inputs (I) and the number of product terms (P) in its PLA file were specified. Each benchmark was run under four different method combinations, with two options for the Polarity (FP and DP) crossed with two options for the Method (D and EO). The evaluation metrics discussed in Section~\ref{sec:evaluation metrics} were included with each method combination. For each benchmark, if the lowest cost was produced by the FP method, it was highlighted in yellow, blue if it was produced by the DP method, and green if both methods produced the same cost. Note that some benchmarks versions could not be run in a feasible amount of time with the exhaustive search methods used in Algorithm EO, so they were omitted from the table. 
\par
\begin{figure}[htbp]
    \centerline{\includegraphics[width=\linewidth]{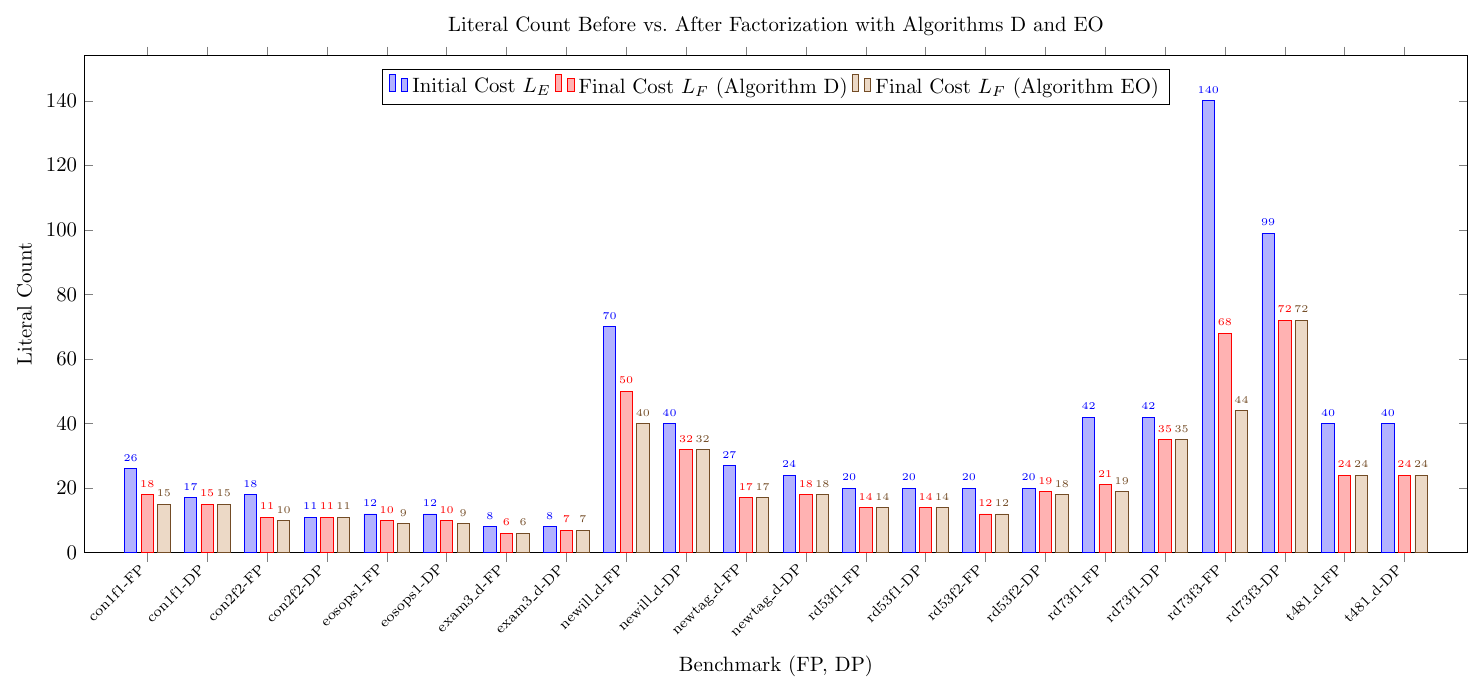}}
    \centerline{\includegraphics[width=\linewidth]{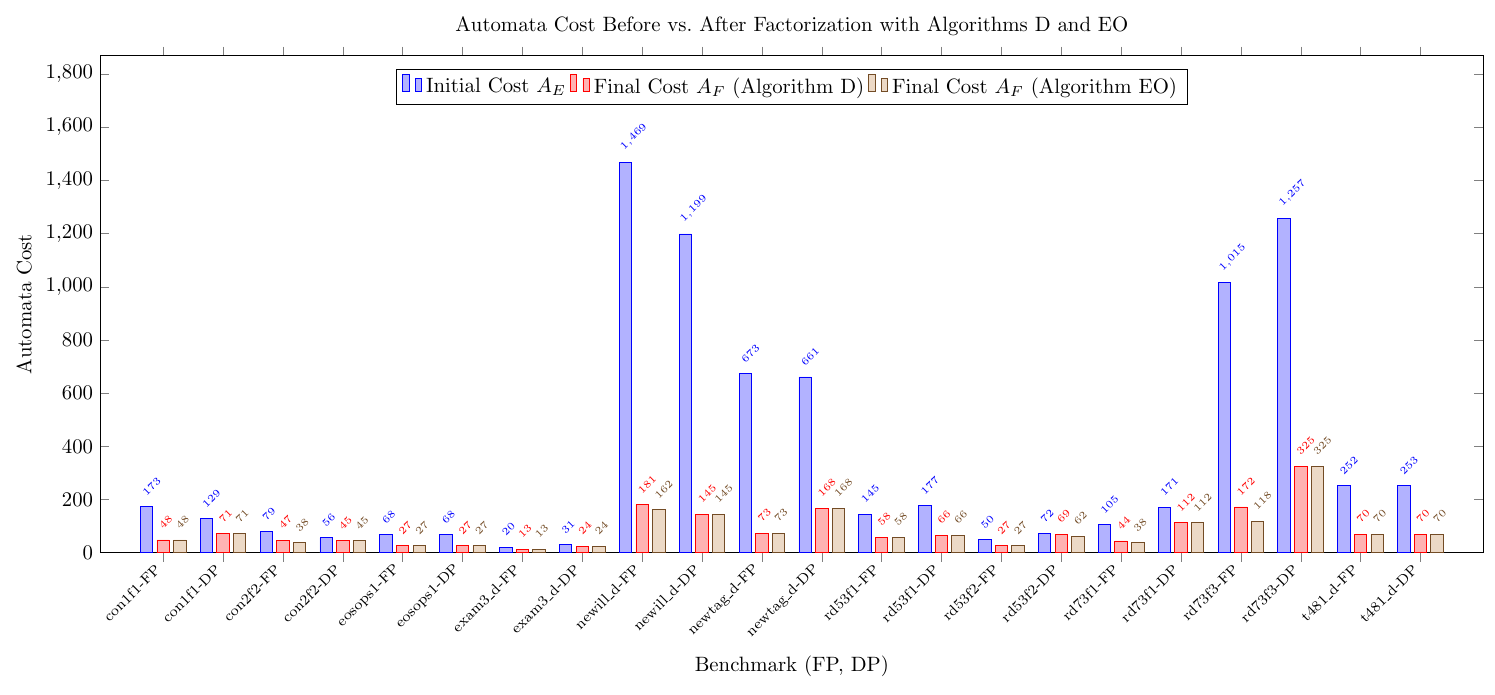}}
    \centerline{\includegraphics[width=\linewidth]{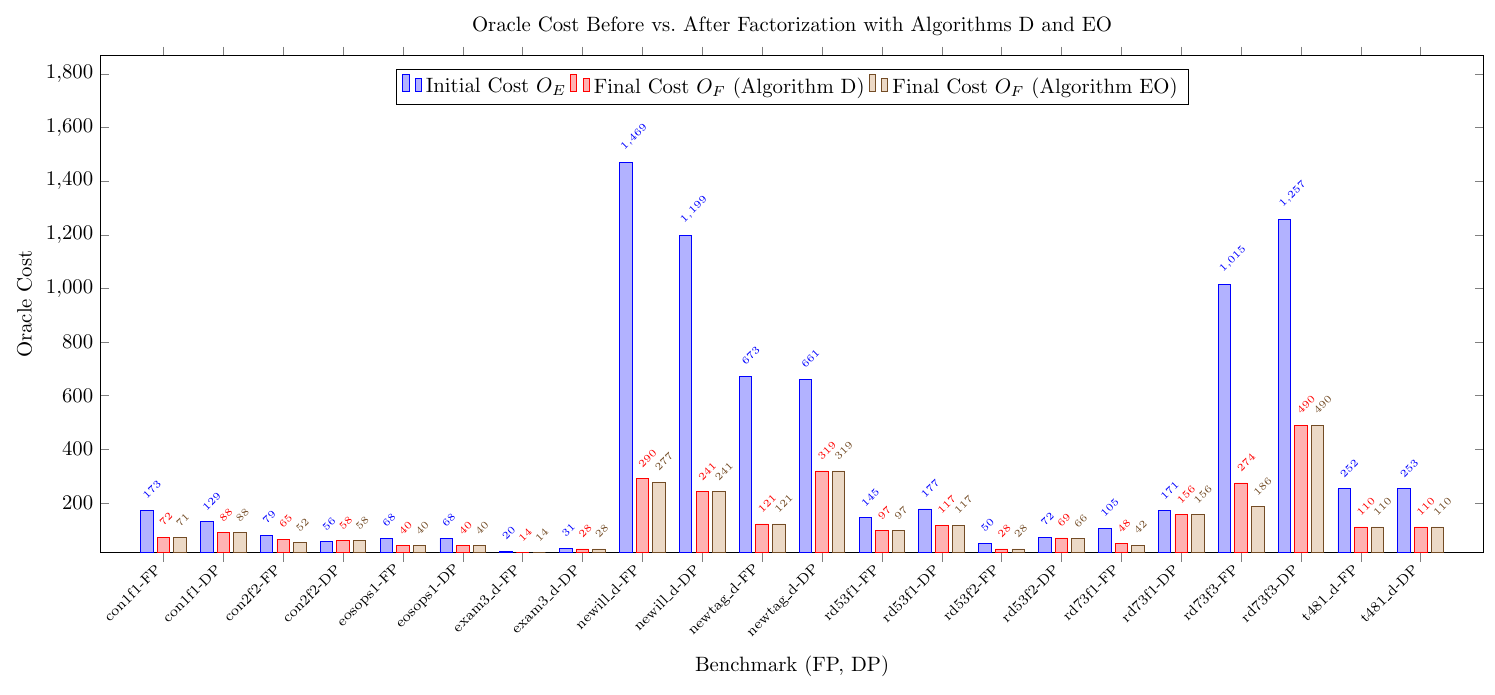}}
    \fcaption{Costs Before and After Factorization}
    \label{fig:bar chart}
\end{figure}

 Figure \ref{fig:bar chart} compares the initial costs with two final costs produced by Algorithms D and EO. Each benchmark was plotted in two instances, one for the FP version and another for DP, which is indicated at the end of each benchmark name. Benchmarks were only included if all four methods combinations were runnable. 
\par
\begin{figure}[htbp]
    \centerline{\includegraphics[width=\linewidth]{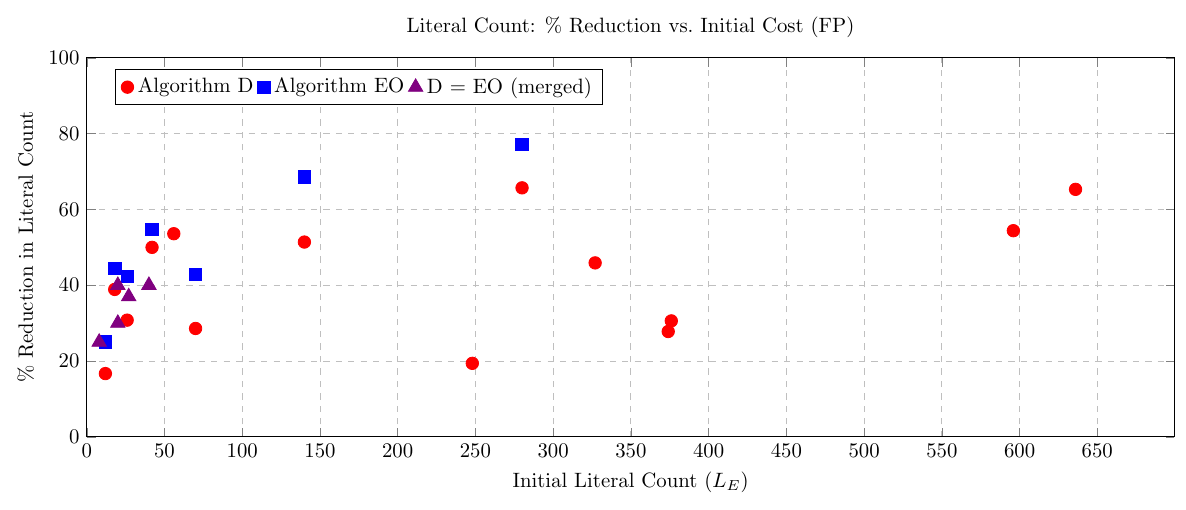}}
    \centerline{\includegraphics[width=\linewidth]{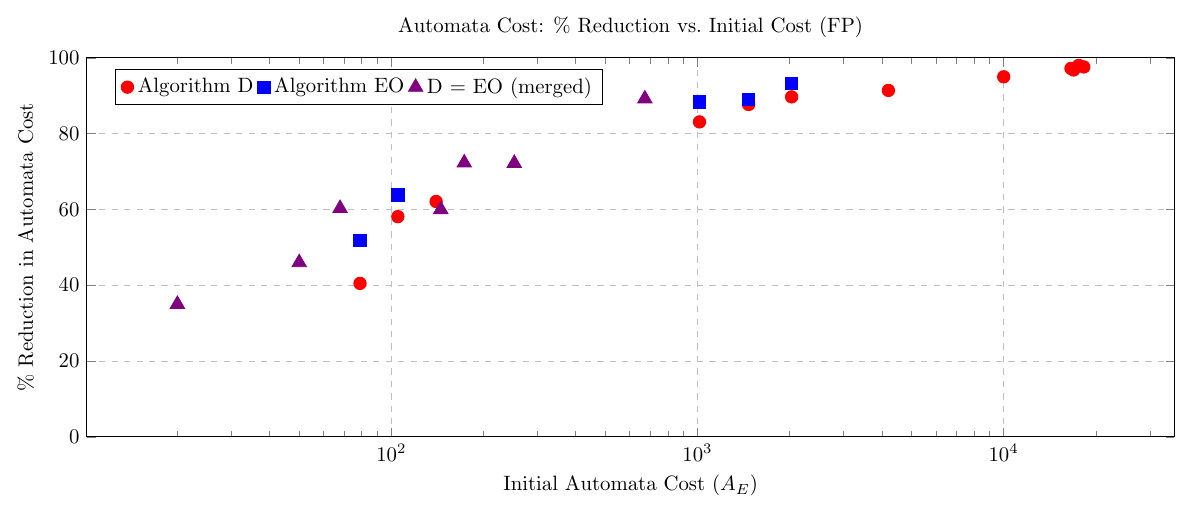}}
    \centerline{\includegraphics[width=\linewidth]{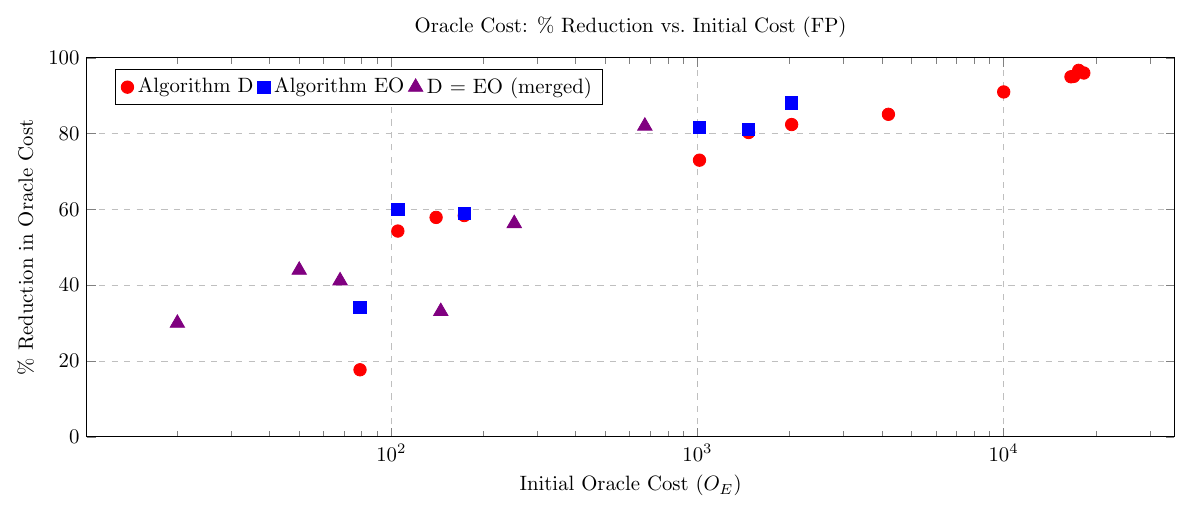}}
    \fcaption{Percentage Reduction in Cost as Initial Cost Increases}
    \label{fig:percentage reduction}
\end{figure}

 Figure \ref{fig:percentage reduction} analyzes the percentage cost reduction based on the initial cost, for only the FP files. For each of the FP version files, two points were plotted: one for the result produced by Algorithm D (red), and another by Algorithm EO (blue). If the result was the same for both algorithms, then the two points were merged (purple). For each method combination, the initial cost was plotted on the $x$-axis, while the percentage cost reduction for that case was plotted on the $y$-axis. Note the logarithmic scale in two of the graphs.
\par Table \ref{tab:FB} compares how different feasibility bounds (FB) affect the results of factorization. In particular, we compared Algorithm EO under an FB of 0.75 vs an FB of 0.9, which allows us to analyze whether the slower algorithm with an FB of 0.75 that checks more rectangles will yield better solutions.

\begin{table*}[htbp]
\scriptsize
\tcaption{Experimental results for expression improvement after applying Algorithms D and EO (FB=0.9) compared to initial expression obtained from EXORCISM-4. Highlighted cells mark the best (lowest) value for $L_F$, $A_F$, or $O_F$ among all method combinations available for that file.}
\label{tab:100}
\setlength{\tabcolsep}{3pt}
\centerline{\renewcommand{\arraystretch}{1.15}
\resizebox{\textwidth}{!}{%
\begin{tabular}{|l|c|c|c|c|c|c|c|c|c|c|c|c|c|c|}
\hline
Benchmark & I & P & Polarity & Method & $L_E$ & $L_F$ & $\Delta L$ & $A_E=O_E$ & $A_F$ & Anc.$_A$ & $\Delta A$ & $O_F$ & Anc.$_O$ & $\Delta O$ \\
\hline
\multirow{4}{*}{con1f1} & \multirow{4}{*}{7} & \multirow{4}{*}{5} & \multirow{2}{*}{FP} & D & 26 & 18 & 30.8\% & 173 & \cellcolor{yellow!60}48 & 3 & 72.3\% & 72 & 1 & 58.4\% \\
 &  &  &  & EO & 26 & \cellcolor{green!20}15 & 42.3\% & 173 & \cellcolor{yellow!60}48 & 3 & 72.3\% & \cellcolor{yellow!60}71 & 1 & 59.0\% \\
\cline{4-15}
 &  &  & \multirow{2}{*}{DP} & D & 17 & \cellcolor{green!20}15 & 11.8\% & 129 & 71 & 2 & 45.0\% & 88 & 1 & 31.8\% \\
 &  &  &  & EO & 17 & \cellcolor{green!20}15 & 11.8\% & 129 & 71 & 2 & 45.0\% & 88 & 1 & 31.8\% \\
\hline
\multirow{4}{*}{con2f2} & \multirow{4}{*}{7} & \multirow{4}{*}{4} & \multirow{2}{*}{FP} & D & 18 & 11 & 38.9\% & 79 & 47 & 4 & 40.5\% & 65 & 1 & 17.7\% \\
 &  &  &  & EO & 18 & \cellcolor{yellow!60}10 & 44.4\% & 79 & \cellcolor{yellow!60}38 & 3 & 51.9\% & \cellcolor{yellow!60}52 & 1 & 34.2\% \\
\cline{4-15}
 &  &  & \multirow{2}{*}{DP} & D & 11 & 11 & 0.0\% & 56 & 45 & 1 & 19.6\% & 58 & 1 & -3.6\% \\
 &  &  &  & EO & 11 & 11 & 0.0\% & 56 & 45 & 1 & 19.6\% & 58 & 1 & -3.6\% \\
\hline
\multirow{4}{*}{eosops1} & \multirow{4}{*}{5} & \multirow{4}{*}{4} & \multirow{2}{*}{FP} & D & 12 & 10 & 16.7\% & 68 & \cellcolor{green!20}27 & 3 & 60.3\% & \cellcolor{green!20}40 & 1 & 41.2\% \\
 &  &  &  & EO & 12 & \cellcolor{green!20}9 & 25.0\% & 68 & \cellcolor{green!20}27 & 3 & 60.3\% & \cellcolor{green!20}40 & 1 & 41.2\% \\
\cline{4-15}
 &  &  & \multirow{2}{*}{DP} & D & 12 & 10 & 16.7\% & 68 & \cellcolor{green!20}27 & 3 & 60.3\% & \cellcolor{green!20}40 & 1 & 41.2\% \\
 &  &  &  & EO & 12 & \cellcolor{green!20}9 & 16.7\% & 68 & \cellcolor{green!20}27 & 3 & 60.3\% & \cellcolor{green!20}40 & 1 & 41.2\% \\
\hline
\multirow{4}{*}{exam3\_d} & \multirow{4}{*}{4} & \multirow{4}{*}{3} & \multirow{2}{*}{FP} & D & 8 & \cellcolor{yellow!60}6 & 25.0\% & 20 & \cellcolor{yellow!60}13 & 1 & 35.0\% & \cellcolor{yellow!60}14 & 0 & 30.0\% \\
 &  &  &  & EO & 8 & \cellcolor{yellow!60}6 & 25.0\% & 20 & \cellcolor{yellow!60}13 & 1 & 35.0\% & \cellcolor{yellow!60}14 & 0 & 30.0\% \\
\cline{4-15}
 &  &  & \multirow{2}{*}{DP} & D & 8 & 7 & 12.5\% & 31 & 24 & 1 & 22.6\% & 28 & 0 & 9.7\% \\
 &  &  &  & EO & 8 & 7 & 12.5\% & 31 & 24 & 1 & 22.6\% & 28 & 0 & 9.7\% \\
\hline
\multirow{4}{*}{newill\_d} & \multirow{4}{*}{8} & \multirow{4}{*}{7} & \multirow{2}{*}{FP} & D & 70 & 50 & 28.6\% & 1469 & 181 & 8 & 87.7\% & 290 & 4 & 80.3\% \\
 &  &  &  & EO & 70 & 40 & 42.9\% & 1469 & 162 & 9 & 89.0\% & 277 & 6 & 81.1\% \\
\cline{4-15}
 &  &  & \multirow{2}{*}{DP} & D & 40 & \cellcolor{cyan!20}32 & 20.0\% & 1199 & \cellcolor{cyan!20}145 & 6 & 87.9\% & \cellcolor{cyan!20}241 & 4 & 79.9\% \\
 &  &  &  & EO & 40 & \cellcolor{cyan!20}32 & 20.0\% & 1199 & \cellcolor{cyan!20}145 & 6 & 87.9\% & \cellcolor{cyan!20}241 & 4 & 79.9\% \\
\hline
\multirow{4}{*}{newtag\_d} & \multirow{4}{*}{8} & \multirow{4}{*}{5} & \multirow{2}{*}{FP} & D & 27 & \cellcolor{yellow!60}17 & 37.0\% & 673 & \cellcolor{yellow!60}73 & 4 & 89.2\% & \cellcolor{yellow!60}121 & 3 & 82.0\% \\
 &  &  &  & EO & 27 & \cellcolor{yellow!60}17 & 37.0\% & 673 & \cellcolor{yellow!60}73 & 4 & 89.2\% & \cellcolor{yellow!60}121 & 3 & 82.0\% \\
\cline{4-15}
 &  &  & \multirow{2}{*}{DP} & D & 24 & 18 & 25.0\% & 661 & 168 & 4 & 74.6\% & 319 & 3 & 51.7\% \\
 &  &  &  & EO & 24 & 18 & 25.0\% & 661 & 168 & 4 & 74.6\% & 319 & 3 & 51.7\% \\
\hline
\multirow{4}{*}{rd53f1} & \multirow{4}{*}{5} & \multirow{4}{*}{5} & \multirow{2}{*}{FP} & D & 20 & \cellcolor{green!20}14 & 30.0\% & 145 & \cellcolor{yellow!60}58 & 5 & 60.0\% & \cellcolor{yellow!60}97 & 3 & 33.1\% \\
 &  &  &  & EO & 20 & \cellcolor{green!20}14 & 30.0\% & 145 & \cellcolor{yellow!60}58 & 5 & 60.0\% & \cellcolor{yellow!60}97 & 3 & 33.1\% \\
\cline{4-15}
 &  &  & \multirow{2}{*}{DP} & D & 20 & \cellcolor{green!20}14 & 30.0\% & 177 & 66 & 5 & 62.7\% & 117 & 5 & 33.9\% \\
 &  &  &  & EO & 20 & \cellcolor{green!20}14 & 30.0\% & 177 & 66 & 5 & 62.7\% & 117 & 5 & 33.9\% \\
\hline
\multirow{4}{*}{rd53f2} & \multirow{4}{*}{5} & \multirow{4}{*}{8} & \multirow{2}{*}{FP} & D & 20 & \cellcolor{yellow!60}12 & 40.0\% & 50 & \cellcolor{yellow!60}27 & 3 & 46.0\% & \cellcolor{yellow!60}28 & 0 & 44.0\% \\
 &  &  &  & EO & 20 & 12 & 40.0\% & 50 & 27 & 3 & 46.0\% & 28 & 0 & 44.0\% \\
\cline{4-15}
 &  &  & \multirow{2}{*}{DP} & D & 20 & 19 & 5.0\% & 72 & 69 & 1 & 4.2\% & 69 & 0 & 4.2\% \\
 &  &  &  & EO & 20 & 18 & 10.0\% & 72 & 62 & 2 & 13.9\% & 66 & 0 & 8.3\% \\
\hline
\multirow{4}{*}{rd73f1} & \multirow{4}{*}{7} & \multirow{4}{*}{15} & \multirow{2}{*}{FP} & D & 42 & 21 & 50.0\% & 105 & 44 & 5 & 58.1\% & 48 & 0 & 54.3\% \\
 &  &  &  & EO & 42 & \cellcolor{yellow!60}19 & 54.8\% & 105 & \cellcolor{yellow!60}38 & 4 & 63.8\% & \cellcolor{yellow!60}42 & 0 & 60.0\% \\
\cline{4-15}
 &  &  & \multirow{2}{*}{DP} & D & 42 & 35 & 16.7\% & 171 & 112 & 5 & 34.5\% & 156 & 2 & 8.8\% \\
 &  &  &  & EO & 42 & 35 & 16.7\% & 171 & 112 & 5 & 34.5\% & 156 & 2 & 8.8\% \\
\hline
\multirow{4}{*}{rd73f3} & \multirow{4}{*}{7} & \multirow{4}{*}{21} & \multirow{2}{*}{FP} & D & 140 & 68 & 51.4\% & 1015 & 172 & 9 & 83.1\% & 274 & 4 & 73.0\% \\
 &  &  &  & EO & 140 & \cellcolor{yellow!60}44 & 68.6\% & 1015 & \cellcolor{yellow!60}118 & 7 & 88.4\% & \cellcolor{yellow!60}186 & 4 & 81.7\% \\
\cline{4-15}
 &  &  & \multirow{2}{*}{DP} & D & 99 & 72 & 27.3\% & 1257 & 325 & 16 & 74.1\% & 490 & 10 & 61.0\% \\
 &  &  &  & EO & 99 & 72 & 27.3\% & 1257 & 325 & 16 & 74.1\% & 490 & 10 & 61.0\% \\
\hline
\multirow{3}{*}{rd84f1} & \multirow{3}{*}{8} & \multirow{3}{*}{21} & \multirow{1}{*}{FP} & D & 56 & \cellcolor{yellow!60}26 & 53.6\% & 140 & 53 & 6 & 62.1\% & 59 & 0 & 57.9\% \\
\cline{4-15}
 &  &  & \multirow{2}{*}{DP} & D & 56 & 44 & 21.4\% & 217 & 127 & 6 & 41.5\% & 183 & 1 & 15.7\% \\
 &  &  &  & EO & 56 & 44 & 21.4\% & 217 & 127 & 6 & 41.5\% & 183 & 1 & 15.7\% \\
\hline
\multirow{3}{*}{rd84f4} & \multirow{3}{*}{8} & \multirow{3}{*}{36} & \multirow{2}{*}{FP} & D & 280 & 96 & 65.7\% & 2030 & 209 & 18 & 89.7\% & 357 & 6 & 82.4\% \\
 &  &  &  & EO & 280 & \cellcolor{yellow!60}64 & 77.1\% & 2030 & \cellcolor{yellow!60}137 & 10 & 93.3\% & \cellcolor{yellow!60}241 & 6 & 88.1\% \\
\cline{4-15}
 &  &  & \multirow{1}{*}{DP} & D & 197 & 150 & 23.9\% & 4440 & 613 & 27 & 86.2\% & 1048 & 17 & 76.4\% \\
\hline
\multirow{4}{*}{t481\_d} & \multirow{4}{*}{16} & \multirow{4}{*}{13} & \multirow{2}{*}{FP} & D & 40 & \cellcolor{green!20}24 & 40.0\% & 252 & \cellcolor{green!20}70 & 4 & 72.2\% & \cellcolor{green!20}110 & 2 & 56.3\% \\
 &  &  &  & EO & 40 & \cellcolor{green!20}24 & 40.0\% & 252 & \cellcolor{green!20}70 & 4 & 72.2\% & \cellcolor{green!20}110 & 2 & 56.3\% \\
\cline{4-15}
 &  &  & \multirow{2}{*}{DP} & D & 40 & \cellcolor{green!20}24 & 40.0\% & 253 & \cellcolor{green!20}70 & 4 & 72.3\% & \cellcolor{green!20}110 & 2 & 56.5\% \\
 &  &  &  & EO & 40 & \cellcolor{green!20}24 & 40.0\% & 253 & \cellcolor{green!20}70 & 4 & 72.3\% & \cellcolor{green!20}110 & 2 & 56.5\% \\
\hline
\multirow{3}{*}{sao2f1} & \multirow{3}{*}{10} & \multirow{3}{*}{10} & FP & D & 248 & 200 & 19.4\% & 10004 & 498 & 18 & 95.0\% & 896 & 10 & 91.0\% \\
\cline{4-15}
 &  &  & \multirow{2}{*}{DP} & D & 80 & \cellcolor{cyan!20}68 & 15.0\% & 5090 & \cellcolor{cyan!20}430 & 8 & 91.6\% & \cellcolor{cyan!20}671 & 7 & 86.8\% \\
 &  &  &  & EO & 80 & \cellcolor{cyan!20}68 & 15.0\% & 5090 & \cellcolor{cyan!20}430 & 8 & 91.6\% & \cellcolor{cyan!20}671 & 7 & 86.8\% \\
\hline
\multirow{3}{*}{sao2f2} & \multirow{3}{*}{10} & \multirow{3}{*}{12} & FP & D & 374 & 270 & 27.8\% & 18276 & \cellcolor{yellow!60}433 & 24 & 97.6\% & \cellcolor{yellow!60}728 & 14 & 96.0\% \\
\cline{4-15}
 &  &  & \multirow{2}{*}{DP} & D & 100 & \cellcolor{cyan!20}67 & 33.0\% & 8156 & 489 & 13 & 94.0\% & 903 & 9 & 88.9\% \\
 &  &  &  & EO & 100 & \cellcolor{cyan!20}67 & 33.0\% & 8156 & 489 & 13 & 94.0\% & 903 & 9 & 88.9\% \\
\hline
\multirow{3}{*}{sao2f3} & \multirow{3}{*}{10} & \multirow{3}{*}{13} & FP & D & 327 & 177 & 45.9\% & 17588 & \cellcolor{yellow!60}353 & 23 & 98.0\% & \cellcolor{yellow!60}575 & 10 & 96.7\% \\
\cline{4-15}
 &  &  & \multirow{2}{*}{DP} & D & 96 & \cellcolor{cyan!20}63 & 34.4\% & 6237 & 408 & 11 & 93.5\% & 744 & 9 & 88.1\% \\
 &  &  &  & EO & 96 & \cellcolor{cyan!20}63 & 34.4\% & 6237 & 408 & 11 & 93.5\% & 744 & 9 & 88.1\% \\
\hline
\multirow{3}{*}{sao2f4} & \multirow{3}{*}{10} & \multirow{3}{*}{11} & FP & D & 376 & 261 & 30.6\% & 16931 & \cellcolor{green!20}539 & 23 & 96.8\% & \cellcolor{yellow!60}834 & 12 & 95.1\% \\
\cline{4-15}
 &  &  & \multirow{2}{*}{DP} & D & 89 & \cellcolor{cyan!20}64 & 28.1\% & 7655 & \cellcolor{green!20}539 & 12 & 93.0\% & 999 & 8 & 86.9\% \\
 &  &  &  & EO & 89 & \cellcolor{cyan!20}64 & 28.1\% & 7655 & \cellcolor{green!20}539 & 12 & 93.0\% & 999 & 8 & 86.9\% \\
\hline
9sym\_d & 9 & 52 & FP & D & 636 & \cellcolor{green!0}221 & 65.3\% & 4204 & \cellcolor{green!0}362 & 27 & 91.4\% & \cellcolor{green!0}625 & 13 & 85.1\% \\
\hline
life\_d & 9 & 50 & FP & D & 596 & \cellcolor{green!0}272 & 54.4\% & 16596 & \cellcolor{green!0}465 & 20 & 97.2\% & \cellcolor{green!0}823 & 9 & 95.0\% \\
\hline
\hline
AVERAGE & - & - & - & - & - & - & 31.12\% & - & - & - & 67.08\% & - & - & 55.09\%\\
\hline
\end{tabular}%
}}
\end{table*}

\begin{table*}[htbp]
\scriptsize
\tcaption{Comparison of feasibility bound (75 vs.\ 90) under FPRM polarity process and EO factorization method.}
\label{tab:FB}
\setlength{\tabcolsep}{3pt}
\centerline{\renewcommand{\arraystretch}{1.3}
\resizebox{\textwidth}{!}{%
\begin{tabular}{|l|c|r|r|r|r|r|c|r|r|c|r|}
\hline
Benchmark & FB & $L_E$ & $L_F$ & $\Delta L$ & $A_E=O_E$ & $A_F$ & Anc.$_A$ & $\Delta A$ & $O_F$ & Anc.$_O$ & $\Delta O$ \\
\hline
\multirow{2}{*}{con1f1} & 75 & 26 & 15 & 42.3\% & 173 & 48 & 3 & 72.3\% & 71 & 1 & 59.0\% \\
\cline{2-12}
 & 90 & 26 & 15 & 42.3\% & 173 & 48 & 3 & 72.3\% & 71 & 1 & 59.0\% \\
\hline\hline
\multirow{2}{*}{con2f2} & 75 & 18 & 10 & 44.4\% & 79 & 38 & 3 & 51.9\% & 52 & 1 & 34.2\% \\
\cline{2-12}
 & 90 & 18 & 10 & 44.4\% & 79 & 38 & 3 & 51.9\% & 52 & 1 & 34.2\% \\
\hline\hline
\multirow{2}{*}{eosops1} & 75 & 12 & 9 & 25.0\% & 68 & 27 & 3 & 60.3\% & 40 & 1 & 41.2\% \\
\cline{2-12}
 & 90 & 12 & 9 & 25.0\% & 68 & 27 & 3 & 60.3\% & 40 & 1 & 41.2\% \\
\hline\hline
\multirow{2}{*}{exam3\_d} & 75 & 8 & 6 & 25.0\% & 20 & 13 & 1 & 35.0\% & 14 & 0 & 30.0\% \\
\cline{2-12}
 & 90 & 8 & 6 & 25.0\% & 20 & 13 & 1 & 35.0\% & 14 & 0 & 30.0\% \\
\hline\hline
\multirow{2}{*}{newill\_d} & 75 & 70 & 40 & 42.9\% & 1469 & 162 & 9 & 89.0\% & 277 & 6 & 81.1\% \\
\cline{2-12}
 & 90 & 70 & 40 & 42.9\% & 1469 & 162 & 9 & 89.0\% & 277 & 6 & 81.1\% \\
\hline\hline
\multirow{2}{*}{newtag\_d} & 75 & 27 & 17 & 37.0\% & 673 & 73 & 4 & 89.2\% & 121 & 3 & 82.0\% \\
\cline{2-12}
 & 90 & 27 & 17 & 37.0\% & 673 & 73 & 4 & 89.2\% & 121 & 3 & 82.0\% \\
\hline\hline
\multirow{2}{*}{rd53f1} & 75 & 20 & 14 & 30.0\% & 145 & 58 & 5 & 60.0\% & 97 & 3 & 33.1\% \\
\cline{2-12}
 & 90 & 20 & 14 & 30.0\% & 145 & 58 & 5 & 60.0\% & 97 & 3 & 33.1\% \\
\hline\hline
\multirow{2}{*}{rd53f2} & 75 & 20 & \cellcolor{green!20}11 & 45.0\% & 50 & \cellcolor{green!20}22 & 2 & 56.0\% & \cellcolor{green!20}24 & 0 & 52.0\% \\
\cline{2-12}
 & 90 & 20 & 12 & 40.0\% & 50 & 27 & 3 & 46.0\% & 28 & 0 & 44.0\% \\
\hline
\end{tabular}
}}
\end{table*}
\newpage
\subsection{Key observations}
Within our data set, both Algorithms D and EO can effectively reduce literal counts by around 20\%-80\%, automata costs by 20\%-90\%, and oracle costs by 20\%-80\% in most of the cases. For all benchmarks, Algorithm EO gives better or equal cost reduction, but the computing complexity is very high. In some cases, Algorithm D gives less optimal solutions than Algorithm EO, but as shown in Fig.~\ref{fig:bar chart}, the relative difference of their final costs is small, which tells us the greedy-based Algorithm D is highly efficient in time complexity and solution output.  
\par In Table \ref{tab:FB}, we see that lowering the FB to 0.75 did not make much a difference compared to a FB of 0.9, except for the benchmark \textit{rd53f2}. 
Lastly, as shown in Fig.~\ref{fig:percentage reduction}, although the percentage of reduction fluctuates when the size of expression is small, it becomes consistently and increasingly higher as the size of expression grows bigger. This shows that our algorithms are much more effective against longer expressions, reducing quantum costs by more than $90$\% in larger examples, which is especially helpful when the initial quantum costs are high.
 
\section{Conclusions}\label{sec:conclusions}
To conclude, the proposed algorithms are very effective at reducing quantum circuit cost, and can reduce literal counts by around 20\%-80\%, automata costs by 20\%-90\%, and oracle costs by 20\%-80\% in most of the cases. In the many of the test cases, the computing efficient Algorithm D shows similar level of reduction as Algorithm EO. Both algorithms demonstrated consistently and significantly higher reduction as the size of expression grows bigger.

\nonumsection{References}

\end{document}